# Exploring the Assessment List for Trustworthy AI in the Context of Advanced Driver-Assistance Systems

Markus Borg, Joshua Bronson
Humanized Autonomy
RISE Research Institutes of Sweden
Lund, Sweden
{firstname.lastname}@ri.se

Linus Christensson, Fredrik Olsson
Dept. of Computer Science
Lund University, Lund, Sweden
{li1768ch-s, fr1284ol-s}@student.lu.se

Olof Lennartsson, Elias Sonnsjö, Hamid Ebabi, Martin Karsberg
Infotiv AB
Gothenburg, Sweden
{firstname.lastname}@infotiv.se

*Abstract*—Artificial Intelligence (AI) is increasingly used in critical applications. Thus, the need for dependable AI systems is rapidly growing. In 2018, the European Commission appointed experts to a High-Level Expert Group on AI (AI-HLEG). AI-HLEG defined Trustworthy AI as 1) lawful, 2) ethical, and 3) robust and specified seven corresponding key requirements. To help development organizations, AI-HLEG recently published the Assessment List for Trustworthy AI (ALTAI). We present an illustrative case study from applying ALTAI to an ongoing development project of an Advanced Driver-Assistance System (ADAS) that relies on Machine Learning (ML). Our experience shows that ALTAI is largely applicable to ADAS development, but specific parts related to human agency and transparency can be disregarded. Moreover, bigger questions related to societal and environmental impact cannot be tackled by an ADAS supplier in isolation. We present how we plan to develop the ADAS to ensure ALTAI-compliance. Finally, we provide three recommendations for the next revision of ALTAI, i.e., life-cycle variants, domain-specific adaptations, and removed redundancy.

## I. INTRODUCTION

AI has surged lately in its uses, relevance, and impact. With that increase, there is a corresponding urgency for ethical reflection on AI [1]. The work of ethics revolves around a central question: how should we live? If AI had little impact on how we may choose to live, if it opened few possibilities, allowed for limited choices, and had highly restricted application areas, it would call for little ethical attention. But, of course, it doesn't. Rather it already affects the way we live in significant ways and holds promises for even further-reaching consequences. Therefore, it is essential that the AI developers, and those who use their products, engage in serious ethical dialogue. This has been increasingly recognized but more needs to be done [2].

The European Commission (EC) communicated an AI strategy in 2018. In the strategy, the EC defines AI as "*systems that display intelligent behavior by analyzing their environment and taking actions – with some degree of autonomy – to achieve specific goals.*" In this paper, we focus on AI that uses Deep Neural Networks (DNN) for supervised Machine Learning (ML), i.e., systems that learn to predict output based on labelled examples.

When the AI Strategy was launched the EC commissioned the High-Level Expert Group on Artificial Intelligence (AI-HLEG); a group of 52 experts representing academia, industry and the civil society. A year later, AI-HLEG published the Ethics Guidelines for Trustworthy AI [3] that puts ethical AI as a key component of trustworthy AI. The guidelines list seven key requirements for trustworthy AI systems.

In 2020, AI-HLEG published the Assessment List for Trustworthy Artificial Intelligence (ALTAI) [4], translating the seven key requirements into several checklists. AI-HLEG claims that "*ALTAI is a practical tool that helps business and organizations to self-assess the trustworthiness of their AI systems under development*." But how applicable is it to an ongoing development project? In this paper we share our experience from applying ALTAI to a contemporary AI system, i.e., the ML-based Advanced Driver-Assistance System (ADAS) SMIRK.

The aim of this study is twofold. First, we want to know how applicable ALTAI is to a typical ML-based ADAS under development. Are the checklists sufficiently concrete to bring value in the automotive development context? Second, we want to use ALTAI to guide design decisions for developing SMIRK as a trustworthy AI product. We use a fictive Company X to support the discussion. Our work is guided by two Research Questions (RQ):

- RQ1. To what extent is ALTAI applicable to ADAS development?
- RQ2. According to ALTAI, how shall SMIRK be designed to meet the criteria for trustworthy AI?

We find that most ALTAI sections are applicable to ADAS development. Exceptions include checklist items related to interactive AI systems and considerations that must be tackled on a systemic level. Extrapolating from ongoing SMIRK development, we provide several recommendations for how Company X could comply with ALTAI. Adherence to existing safety and security engineering standards goes a long way, but it must be complemented by ML best practices such as pipelines architectures, data version control, and experiment tracking.

The rest of this paper is organized as follows. Section 2 introduces trustworthy AI according to AI-HLEG and the fundamentals of safe automotive AI. Section 3 presents the case, i.e., Company X and SMIRK. Section 4 describes our research method and Section 5 discusses our results. Finally, Section 6 summarizes our findings and presents recommendations for a future ALTAI revision.

## II. Background

This section introduces the European Commission's work on trustworthy AI and a general introduction to automotive software safety in the AI era.

### A. Trustworthy AI in the European Union

In the 2019 publication of the AI-HLEG "Ethics Guidelines for Trustworthy AI" they state that AI must be: Lawful – complying with all applicable laws and regulations; Ethical – ensuring adherence to ethical principles and values; Robust – both from a technical and social perspective.

AI-HLEG presents guidelines as a framework for achieving trustworthy AI. However, the guidelines do not address the lawful component, so it shouldn't be considered legal advice. Instead, the guidelines specify seven key requirements for trustworthy AI – we refer to them in brackets, e.g., **[REQX]**.

**[REQ1] Human agency and oversight**. An AI system shall "empower human beings, allowing them to make informed decisions and fostering their fundamental rights." Oversight mechanisms must allow humans to inspect the decisions made. Oversight can be achieved through three mechanisms: 1) human-in-the-loop (intervention in every decision cycle), 2) human-on-the-loop (intervention during design and monitoring), and 3) human-in-command (wide oversight with ability to deactivate AI).

**[REQ2] Technical robustness and safety**. An AI system shall be "resilient and secure." Prevention of harm must permeate the design of the system. Security is fundamental to protect against antagonistic attacks. Safety is essential to ensure operation without harming people or the environment. Furthermore, AI systems must be sufficiently accurate and reliable.

**[REQ3] Privacy and data governance**. An AI system shall "ensure full respect for privacy and data protection." This requires adequate data management policies and data governance mechanisms to ensure the quality and integrity of the data. Moreover, a secure system for data access control must exist.

**[REQ4] Transparency**. The AI system shall be open about the underlying "data, system and AI business models." Humans must understand when they are interacting with an AI system. Output must be explained, and users shall be informed of inherent capabilities and limitations. Traceability mechanisms are key components to achieve transparency.

**[REQ5] Diversity, non-discrimination and fairness**. An AI system shall be "accessible to all, regardless of any disability, and involve relevant stakeholders throughout their entire life circle." Unfair bias can have severe negative implications and must be avoided. Diversity shall be promoted; thus, systems must be designed for a broad demographic profile.

**[REQ6] Societal and environmental well-being**. An AI system shall **"**benefit all human beings, including future generations." This requirement covers how AI affects the environment and society at large. Example considerations include sustainability and how systems can have societal impact.

**[REQ7] Accountability**. An AI system shall "ensure responsibility and accountability for AI systems and their outcomes." Mechanisms must be in place to enable auditability, i.e., the assessment of algorithms, data and design processes. Accountability relates to risk management and mitigation.

### B. AI in Safety-critical Automotive Software

Modern cars are software-intensive systems. Some of the automotive software controls safety-critical functions and must thus be developed according to rigorous safety engineering principles. Safety is defined as a negative concept, i.e., "*absence of unacceptable risk.*" Considerable effort is required to develop a system's safety case, i.e., a structured argumentation backed by evidence. In the automotive domain, functional safety is covered by the standard ISO 26262, an automotive adaptation of the generic IEC 61508 standard.

In the last decade, DNNs revolutionized computer vision applications. In the automotive context, DNNs enabled substantially more accurate vehicular perception. However, embedding DNNs in safety-critical software equals a paradigm shift. No longer is the functional behavior explicitly coded by a developer, instead enormous amounts of labelled data are used to train a mapping function between input and output. ISO 26262 does not fit the ML paradigm [5] as prescribed practices such as comprehensive code coverage testing and code reviews miss the target.

In 2019, ISO/PAS 21148 was published, a stepping-stone toward a new standard. ISO 21448 is intended to complement functional safety for automotive systems that rely on ML. A key concept in ISO 21448 (and other standards) is the Operational Design Domain (ODD), i.e., the specific operating context in which a function is designed to perform. An ODD specifies aspects such as roadway types, speed ranges, weather conditions, traffic densities and the presence of non-road objects.

Automotive engineering relies heavily on standards. Apart from the previously mentioned examples, this paper will refer to the following: 1) SAE J3016 Taxonomy and Definitions for Terms Related to On-Road Motor Vehicle Automated Driving Systems 2) ISO/TS 16949 Automotive Quality Management, 3) ANSI/UL 4600 Standard for Safety for the Evaluation of Autonomous Products, 4) ISO/SAE DIS 21434 Road vehicles - Cybersecurity engineering, 5) ISO/IEC 27701 Security Techniques for Privacy Information Management - Requirements and Guidelines, and 6) ISO/TS 8000-150 Data quality - Part 150: Master Data: Quality Management Framework.

## III. Case Description

This section describes Company X and SMIRK.

### A. Company X – A Fictive Automotive Tier 1 Supplier

The automotive industry is a highly complex and multi-tiered network of suppliers. Fig. 1 displays the general three-tier setup of the automotive supply chain. The y-axis represents value refinement with the tip of the iceberg containing the car manufacturers, i.e., the Original Equipment Manufacturers (OEM). At the bottom is Tier 3, suppliers of raw materials such as metal and plastic as well as basic, standardized parts. Tier 2 suppliers provide components, but are not necessarily restricted to the automotive industry. These suppliers typically do not produce components that are automotive-grade, and thus do not sell directly to the OEMs. From Tier 1 and upwards, the automotive-specific quality and safety standards enter the picture. Tier 1 suppliers provide modules and (sub-)systems directly to the OEMs. These suppliers can sell their products to several OEMs, but tend to have a tight connection to only one or two. Suppliers of ADAS are examples of Tier 1 actors.

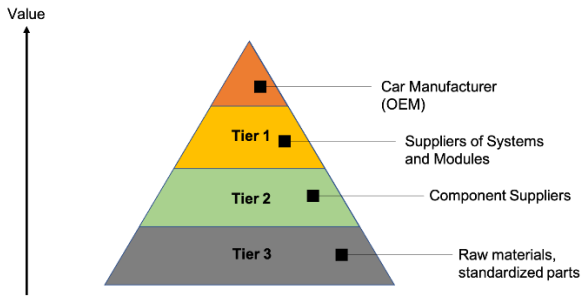

Figure 1: The three-tier automotive supply chain.

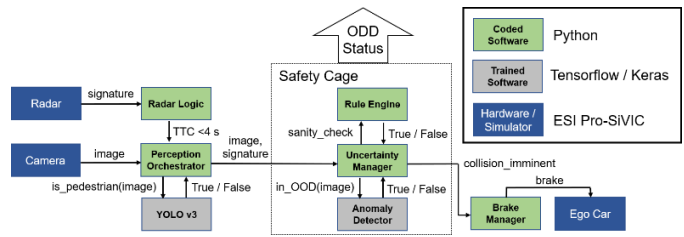

Figure 2: Logical view of the SMIRK architecture.

The automotive industry is moving toward more software-intensive systems and relies increasingly on variability to diversify products on the market. Consequently, Tier 1 suppliers are playing a more important role in the industry [6]. Due to OEMs' growing dependence on Tier 1 suppliers, it is essential they show that any AI-enabled system is trustworthy. We expect assessments such as ALTAI to become a prerequisite for business with OEMs.

Company X is a fictive Tier 1 supplier specializing in active safety solutions and currently developing a bespoke ADAS tailored for a major OEM. The ADAS, called SMIRK, is described in Section 2.B.

### B. SMIRK – The Case ADAS under Study

The case system under study is SMIRK, an Open-Source Software (OSS) ML-based ADAS under development. SMIRK is a research prototype that provides pedestrian emergency braking. To ensure industrial relevance, SMIRK builds on the reference architecture from PeVi, an ADAS studied in previous work [7]. The SMIRK development adheres to development practices mandated by the candidate standard ISO/PAS 21448.

PeVi uses a radar sensor and a camera to detect pedestrians in "warning areas" in front of the vehicle and alerts the driver in case of an imminent collision. SMIRK will provide an equally capable OSS ADAS, combining Python source code and a trained DNN for object detection. In contrast to PeVi, SMIRK will also initiate emergency braking. Thus, SMIRK will constitute safety-critical driving automation on SAE Level 2.

Fig. 2 shows the (simplified) logical view of the SMIRK architecture. The view illustrates an ADAS combining software implemented in Python, a trained object detection model (the DNN YOLOv3), and two hardware sensors. Analogous to PeVi, SMIRK will first rely on the radar sensor to detect potential collisions with external objects. If the radar signature corresponds to an imminent collision between the ego car and another object, i.e., a low value (<4 seconds) for the Time-To-Collision (TTC), SMIRK will dispatch an image from the camera to the trained YOLO. If the model predicts the presence of a pedestrian, SMIRK will avoid a collision by emergency braking.

Output from the object detection model corresponds to one of four outcomes: 1) True Positive (TP, detecting a pedestrian that is present in front of the car), True Negative (TN, no detection and there is no pedestrian), False Positive (FP, detecting a pedestrian where no one is present) and False Negative (FN, missing a pedestrian). While both FPs and FNs need to be kept at a minimum, they are not equally bad for SMIRK – illustrating a common trade-off in ADAS development.

Based on a SMIRK Hazard Analysis and Risk Assessment workshop, we conclude that the main hazards involved relate to FPs ("breaking for ghosts") rather than FNs (not breaking for a pedestrian). A human driver can break if SMIRK fails to detect a pedestrian, but incorrect emergency braking might lead to rear-end collisions. Appropriate safety mechanisms must be implemented to alleviate the corresponding risks.

SMIRK uses a safety cage mechanism to mitigate the hazards involved in false negatives. The safety cage acts as a supervisor, i.e., an uncertainty estimator, rejecting input that does not resemble the DNN's training data [8]. Input images are assessed by a combination of trained (unsupervised) anomaly detection and rule-based heuristics, e.g., violations of the laws of physics. If the input image appears highly anomalous, the safety cage will reject the input, i.e., the DNN output won't be trusted, and SMIRK will request a human handover.

The virtual prototyping solution ESI Pro-SiVIC is used to develop SMIRK. The simulated environment allows advanced dynamic modeling of vehicles and pedestrians as well as high-fidelity sensor models of cameras and radar. We also use the tool to create synthetic training data for the DNN.

To support the later discussion, we define a roadmap with four versions of SMIRK. The versions represent evolutionary development of SMIRK, but some fundamental variation points lead to different outcomes in the ALTAI assessment.

**SMIRK-MVP**: A minimum viable product with a highly restricted ODD (straight road, perfect conditions, and only a single pedestrian). A DNN exclusively trained on synthetic data.

**SMIRK-Local (v1.0)**: The first major release that meets the bespoke requirements by the OEM: inclusive ODD (elevation, weather, curves, scenery, and multiple pedestrians), mixed synthetic and real data training, and no needed network connection.

**SMIRK-Remote (v2.0)**: Same ODD and mixed training data but connected to the cloud with continuous data collection and nightly uploads. Company X uses data to train improved DNNs and pushes regular updates to the fleet of cars.

**SMIRK-Federated (v3.0)**: Adds online learning to adapt to the specific car, retraining DNNs nightly. Uses federated learning as a privacy-preserving approach to share DNN updates with the rest of the fleet via the cloud (as described by Du *et al.* [9]).

## IV. METHOD

We conducted an illustrative case study by using ALTAI on SMIRK. The diverse team of researchers provided multiple perspectives from academia (industrial economics, software engineering, and ethics) and industry (automotive engineering, software testing, and V&V). Before the study, all authors read the AI-HLEG's Ethics Guidelines for Trustworthy AI and took part in workshops within the SMIRK development project.

Fig. 3 shows an overview of the research method behind the results presented in this paper. First, authors 3-4 studied ALTAI and the SMIRK documentation. The activity concluded with a draft assessment. Second, authors 1-2 studied ALTAI and iteratively refined the assessment together with authors 3-4. This intermediate step involved an explicit ethical perspective. Third, authors 5-8 performed an assessment of the work from an industrial perspective. This step resulted in a validated assessment. Finally, all authors together collected key findings in this paper.

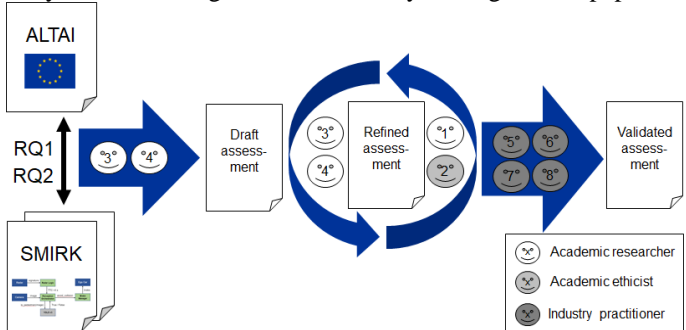

Figure 3: Overview of the research method. Nose numbers map to order of the authors.

## V. RESULTS AND DISCUSSION

This section is organized according to the seven key requirements for trustworthy AI. To appreciate this section, we recommend the reader to simultaneously have our supporting document[1] available for cross-checking. Each subsection follows the same structure. First, we discuss which of the ALTAI Questions (Q) that are not applicable (N/A) to an ADAS. Second, we report our findings from using ALTAI on SMIRK. When referring to specific ALTAI Qs, we use the following format in parentheses:

(**<KeyRequirement>.<Q>.<SubQ>**).

### A. [REQ1] Human Agency and Oversight

**[REQ1]** consists of **Human Agency and Autonomy** (**HA&A Q1.1-1.6**) and **Human Oversight** (**HO Q1.7-1.11**).

*1) ALTAI Applicability for ADAS*

---
[1] https://tinyurl.com/ALTAI-Qs

Several ALTAI questions related to **[REQ1]** are not applicable to an ADAS such as SMIRK. Many ADAS do not directly interact with human users. Instead, the human driver primarily interacts with the ego car that consists of many subsystems. To allow a smooth user interaction, the human should not need to bother with independent messages from tens of subsystems. The OEM must integrate the user interaction into a consistent user interface, e.g., through dashboard lights. We conclude that **Q1.2-1.2.1** are N/A. In the same vein, **Q1.5-1.6.3** are N/A as they address the risk of humans developing unhealthy relationships with the system – an ADAS shall not offer any social interaction.

We find that most of the Qs related to **[REQ1]** are applicable to ADAS. Human agency and oversight are important aspects in this context. On the other hand, several questions specifically target AI systems with direct human interaction. Although an ADAS can provide information that an OEM can propagate to a human driver, ADAS primarily interacts with other subsystems in the ego car. In a future revision of ALTAI, we recommend moving Qs related to AI systems that directly interact with humans to a separate subsection.

*2) SMIRK ALTAI Assessment*

**HA&A.** To avoid over-reliance on SMIRK (**Q1.3.1**), the driver must know when SMIRK is within its ODD or not. This could be communicated through a dashboard light, i.e., indicating whether SMIRK is ready to intervene. Another option would be to use audio signals, but such cues risk annoying drivers. However, the OEM must make this design decision. SMIRK uses the safety cage to determine the ODD status during operation and propagates the information to higher architectural levels in the ego car, as depicted by the "ODD Status" arrow in Fig. 2).

Another related communication aspect is informing the user when the autobrake has been activated because of pedestrian detection. This is important to avoid confusion as to why the ego car suddenly actuated the braking system. A key design concern for SMIRK is avoiding unintended interference with the driver (**Q1.4**). SMIRK's safety cage is the main mechanism for achieving this, i.e., a construct aiming at reducing the number of FPs.

**HO.** SMIRK does not use any input or approval from the user in its decision making, i.e., SMIRK has a human-on-the-loop rather than a human-in-the-loop oversight structure (**Q1.7**). Due to the human-on-the-loop construct, and the absence of procedures to abort SMIRK's momentaneous decision making (**Q1.10**), it is crucial that the developers at Company X receive proper training on how to exercise oversight (**Q1.8**).

Detection and response to the unintentional behavior of SMIRK (**Q1.9**) differs between the different versions. Engineers can look into the decisions that are made by SMIRK-Remote and SMIRK-Federated as they continuously store camera images, radar data, SMIRK output, and the state of the ego car for a set period of time. SMIRK-Remote transmits the information to Company X to allow *post hoc* analyses of anomalous situations. SMIRK-Federated both transmits the information and implements continuous learning to updates the local DNN on a nightly basis. Learning from FNs (missed pedestrians) can be based on a comparison with the driver's brake action or detected collisions. Learning from FPs (unmotivated braking), however, requires an explicit feedback mechanism involving the driver, e.g., a feedback button or voice control. On the other hand, SMIRK- MVP and SMIRK-Local do not have any such features as they are not connected to any communication networks.

*B. [REQ2] Technical Robustness and Safety*

**[REQ2]** contains four subsections: **Resilience to Attack and Security (RA&S Q2.1-2.6)**, **General Safety (GS Q2.7-2.11)**, **Accuracy (A Q2.12-2.16)**, and **Reliability, Fall-back plans and Reproducibility (RF&R Q2.17-2.21)**.

*1) ALTAI Applicability for ADAS*

**[REQ2]** is the most comprehensive requirement proposed by AI-HLEG, covering several fundamental engineering concerns. On one hand, it is an advantage that safety and security are combined in the same key requirement – co-engineering these quality attributes is a recommended practice in automotive engineering [10]. On the other hand, it makes the ALTAI assessment unbalanced. A possible solution could be to split the very technical [**REQ2**] into "Accuracy and Robustness" and "Safety and Security".

As both technical robustness and safety are cornerstones of ADAS, all questions are relevant. The [**REQ2**] questions are generally very broad and, in line with common practice when developing assurance cases [11], it is left to the development organization to argue that sufficient technical robustness and safety measures have been taken. We argue that this ALTAI section would benefit from explicit references to existing security and safety standards, both of general and domain-specific nature. Finally, we find that the detailed Qs on antagonistic attacks in [**Q2.3-2.3.2**] are somewhat redundant as [**Q2.2**] already covers security certification.

*2) SMIRK ALTAI Assessment*

**RA&S.** SMIRK could be exposed to technical faults or attacks, resulting in threat to the driver and others (**Q2.1**). For example, a sudden, unprovoked autobrake on a highway at high speed could be disastrous. Integrity, robustness and overall security must be prioritized throughout SMIRK's entire life cycle, from the design phase until the system is phased out. SMIRK must be certified for automotive cybersecurity according to ISO/SAE 21434 (**Q2.2**), including third-party security audits by ML experts (data poisoning and model inversion). However, as SMIRK will rely on the overall network communication in the vehicle, the OEM will lead the major security effort.

Company X must establish a vulnerability management process and share information openly with the OEM and the ecosystem [12] (**Q2.6**). Furthermore, Company X must provide security patches as long as SMIRK systems are operating (**Q2.6.1**). We argue that the security certification subsumes penetration testing (**Q2.3-2.5**).

**GS.** SMIRK is a safety-critical automotive system and must be treated accordingly. Company X must define risk, risk metrics, and risk levels according to ISO 26262 and acquire an appropriate safety certification for SMIRK (**Q2.7**). ISO 26262 defines risk as a composition of severity, probability of exposure, and driver controllability – resulting in an Automotive Safety Integrity Level (ASIL). ASIL is a domain-specific adaptation of the SIL in IEC 61508, defined as a target level of risk-reduction. However, as discussed in Section 2B, ISO 26262 must be complemented by ISO 21448 for automotive systems that rely on ML. ISO 26262 and ISO 21448 certification, possibly using the prompts presented in ANSI/UL 4600 as a checklist, subsume the remaining GS Qs (**Q2.8-Q2.11**).

**Accuracy.** Before deploying SMIRK, it will be pivotal to ensure a highly accurate DNN, especially when the ADAS is inside its ODD (**Q2.12**). In order to ensure high accuracy, the dataset (training, validation, and test) used in the SMIRK development must be targeted by a quality assurance process (**Q2.13**). Company X will also involve a third party to perform independent verification & validation, i.e., a party that is not involved in the product development. Processes and documentation must be implemented to ensure that the accuracy of SMIRK is continuously measured (including the FP/FN trade-off described in Section 3.B) and communicated to the OEM (**Q2.13, Q2.16**).

SMIRK-Remote and SMIRK-Federated continuously gather new data during operations, which requires automated quality assurance both on the vehicle and in the cloud. On the vehicle, it is essential that sufficient meta-data is collected (e.g., weather conditions and geo-position) to allow fine-granular data versioning. Reliable meta-data is also required if Company X decides to extend the ODD in future generations of SMIRK. As the data is fundamental in ML-based systems, all information about the operational context from which it was collected must be stored.

The CACE principle is a well-known issue in ML, i.e., "Changing Anything Changes Everything" [13]. ML entangles the information propagation through DNNs so that nothing is ever truly independent. Thus, it is currently impossible to isolate improvements. Retraining a DNN with some additional training images risks invalidating the dataset as a whole, resulting in inaccurate predictions (**Q2.15**). An example of such a situation could be retraining on images of holograms for speed control as presented in ISO 21448.

ML development must be highly iterative, and careful experiment tracking is essential as well as rigorous regression testing (**Q2.14-2.15**). For each newly trained model, Company X runs a series of automated statistical test cases to detect any model deterioration. SMIRK v3.0 runs these activities both on the vehicle and overnight in the cloud.

**RF&R.** Company X must establish processes and documentation to continuously verify and validate the reliability and reproducibility of SMIRK (**Q2.17**), including comprehensive logging (**Q2.18**). As mentioned in the accuracy subsection, data versioning and experiment tracking is fundamental to reproducibility. Company X uses a contemporary ML tech stack to support these aspects, i.e., DVC (dvc.org) for data version control and CML (cml.dev) for experiment tracking in a continuous integration context based on GitHub Actions.

The safety cage mechanism in SMIRK, addresses results with low confidence scores (**Q2.20**). When the safety cage rejects input, the failsafe fallback plan is to request a handover to the driver (**Q2.19**). Testing of the fallback plan is covered by the safety certification process. The safety cage is even more important to SMIRK v3.0, as it uses online learning (**Q2.21**).

*C. [REQ3] Privacy and Data Governance*

**[REQ3]** contains two subsections: **Privacy** (**Q3.1-3.2**) and **Data Governance** (**DG Q3.3-3.6**).

*1) ALTAI Applicability for ADAS*

There is a recognized contradiction in ML development between collecting enormous amounts of training data and preserving the privacy of humans that might be included in the data. The introduction of GDPR exacerbated the issue in the EU [14]. In June 2020, the European Parliamentary Research Service published a report on the impact of GDPR on AI [15]. The report acknowledges the tension, but claims that AI is indeed compatible with GDPR. Approaches to tackle GDPR compliance might include anonymization or pseudonymization of data and safeguards for inferences based on personal data. A data governance process, supported by the data versioning discussed under **[REQ2]**, will be necessary to manage user consent and data purpose. Regarding the feared "right to be forgotten" clause of GDPR, i.e., that EU residents can request an organization to erase all personal data without undue delay, the report suggests that users should have the right to be removed from the dataset. On the other hand, ML models trained using that data do not need be erased - the embedded data is not personal after training.

**[REQ3]** is highly relevant to an ML-based ADAS. ADAS use ML to enable vehicular perception, i.e., understanding the surrounding traffic. The underlying ML models are trained on input collected from sensors. Especially cameras are sensitive, as the collected data inevitably contains personal data such as human faces. Moreover, images showing clothing or background buildings might reveal sensitive information such as political and religious affiliations. We find all Qs relevant, but propose two improvements for the next ALTAI revision: 1) make ALTAI more actionable by recommending mechanisms to tackle the listed considerations, and 2) explicitly suggest data governance standards in **Q3.6**.

*2) SMIRK ALTAI Assessment*

**Privacy.** SMIRK MVP is exclusively trained on synthetic data, thus there are no privacy concerns to consider. For the other versions data protection is central (**Q3.1**). Together with the OEM, and probably other ADAS suppliers, Company X must establish a feedback channel that allows users to flag potential privacy issues via the OEM (**Q3.2**). SMIRK securely stores data in an encrypted format both on the vehicle and in the cloud and all network communication is encrypted. Furthermore, SMIRK v3.0 uses privacy-preserving federated learning. Finally, Company X uses automatic face anonymization [16], which also supports fairness by masking protected attributes (see Sec. 5.E).

**DG.** Data governance is crucial for SMIRK v1.0, v2.0, and v3.0 (**Q3.3**). Company X is based in the EU and complies with GDPR (**Q3.4**). Example measures that are put in place include a dedicated Data Protection Officer and GDPR training for the employees. Only qualified personnel who underwent the appropriate level of training have access to the sensitive dataset. Also, Company X has implemented mechanisms to log when, where, by whom and for what purpose the dataset is accessed. Fine-granular data version control with DVC and the feedback channel (Q3.2) enable the GDPR right to be forgotten (**Q3.4.1**).

Company X adheres to two additional data standards to further support privacy and data governance. As GDPR requirements are expressed in a way that will not turn obsolete anytime soon, it largely tells an organization what to do but not how to do it. To remedy this, Company X is certified according to ISO/IEC 27701, i.e., implementing several mechanisms to ensure data privacy. Finally, Company X follows part 150 of ISO/TS 8000, i.e., the data quality management framework.

*D. [REQ4] Transparency*

**[REQ4]** consists of **Traceability** (**Q4.1-4.1.5**), **Explainability** (**Q4.2-4.3**), and **Communication** (**Q4.4-4.5.1**).

*1) ALTAI Applicability for ADAS*

**[REQ4]** covers the overall interpretability of the AI system and its output. Software traceability in a system that relies on a DNN is difficult [17]. Data lineage is a closely related concept in ML, describing what data originates from where and how it has been processed. Both traceability and lineage are necessary to be able to backtrack from the output of an ML model to instances of the training data. Backtracking is fundamental in root cause analysis as part of post-accident investigations. The ALTAI traceability subsection is thus highly relevant to ADAS.

The explainability subsection is also relevant to ADAS. With several ADAS operating simultaneously in a modern car, each individual ADAS cannot share its decision rationales with the driver at each instant. However, an ADAS must undoubtedly be able to explain its output when explicitly requested. Stakeholders are not limited to drivers and the OEM, but also include the legal system, insurance companies, and government agencies.

The communication subsection is not applicable to a typical ADAS, i.e., **Q4.4-4.5.3** are N/A. An ADAS is not an interactive AI system that communicates directly with the driver, and there is no risk that the driver mistakes the system for another human. Also, communicating the benefits and limitations of increased vehicle automation in general must be done by the OEM.

*2) SMIRK ALTAI Assessment*

As OSS, SMIRK is an unusually transparent ADAS. Still, despite the openness, the corresponding Qs must be discussed.

**Traceability.** Company X uses a data pipeline to automate a sequence of various processing steps. Together with data version control, the pipeline architecture ensures data lineage (**Q4.1**). The first steps of the pipeline cover data validation (**Q4.1.1**), i.e., basic sanity checks to reject low quality data (such as data collected from blocked or broken sensors). Validated data continues to standard computer vision preprocessing steps, e.g., image resizing, noise removal, and smoothing of edges. Preprocessed data carry on to data augmentation, enabling the training of more robust ML models through image rotations, brightness changes, and modified weather conditions. In SMIRK v3.0, the data pipeline is deployed on the vehicle to allow federated learning.

During operation, SMIRK logs all decisions and their corresponding sensor input, i.e., camera images and radar signatures (**Q4.1.5**). In a *post hoc* analysis, the logged information can be used to reproduce the output from the ML model and which training data were decisive in the prediction (**Q4.1.2**). For SMIRK, the traceability must distinguish between input rejected by the safety cage and incorrect predictions by the DNN.

**Explainability.** Using the described traceability solution, Company X will be ready to explain the output of specific SMIRK decisions whenever requested by the OEM (**Q4.2**). The OEM will conduct standard customer satisfaction surveys, including the perception of the integrated ADAS (**Q4.3**).

**Communication.** Since SMIRK does not communicate with the driver, Company X will only support the OEM in its pedagogical duties of explaining the ADAS toward the customers.

*E. [REQ5] Diversity, Non-discrimination and Fairness*

**[REQ5]** contains three subsections: **Avoidance of Unfair Bias** (**AUB Q5.1-5.5.4**), **Accessibility and Universal Design** (**AUD Q5.6-5.9.3**), and **Stakeholder Participation** (**SP Q5.10**).

*1) ALTAI Applicability for ADAS*

**[REQ5]** is the second-longest part of ALTAI, primarily discussing diversity and fairness. The main focus of the Qs is on measuring and minimizing bias and developing fairness supporting procedures. Unfair bias and freedom from discrimination are important aspects to tackle in the requirements engineering of an ML-based ADAS [18]. The developers must identify protected attributes (e.g., gender, age, and ethnicity) and use available tools to balance the training data accordingly. AUB is indeed an applicable ALTAI subsection. A major issue with ALTAI is that **Q5.5** requests an explicit definition of fairness without proposing any alternatives. Fairness is a non-trivial concept, and a revised ALTAI should propose domain-specific recommendations to help development organizations.

We argue that the AUD subsection is irrelevant to ADAS development. It is certainly important that drivers with special needs receive critical feedback from an ADAS, but it is up to the OEM to design user interfaces that are appropriate for various types of drivers. As a typical ADAS does not directly communicate with the driver, we consider **Q5.6-5.9.3** as N/A.

*2) SMIRK ALTAI Assessment*

**AUB.** As the object detection model (YOLO) of SMIRK is ML-based, it will never perform better than its training data. Developing a data management strategy that minimized unfair bias is a key activity. Company X use a multi-faceted approach (incl. stakeholder participation) to develop a strategy with 1) data collection and 2) a corresponding V&V plan ensuring the unbiased collection of training data with protected characteristics (**Q5.1**).

The data management strategy considers a broad demographic profile based on statistically average cross sampling of the EU population. Furthermore, Company X use Section 7.3.2 of UL 4600 as a checklist of user groups that must be included to produce a fair representation (**Q5.2**). To test the ML model for bias, Company X uses the AI Fairness 360 Toolkit provided by IBM Research (**Q5.2.1-5.2.3**).

To mitigate implicit bias, all Company X employees participate in internal fairness training (**Q5.3**). The training recognizes that fairness is complex and depends on context and culture, with no single definition capturing all aspects of fairness. In line with the definition of safety, and partly inspired by IBM Research [19], Company X defines fairness as a negative quality: "fairness is absence of unwanted bias that places privileged groups at a systematic advantage and unprivileged groups at a systematic disadvantage" (**Q5.5**). Finally, just as users can flag privacy issues via the OEM, the same feedback channel can be used to report detected bias (**Q5.4**).

**AUD.** N/A, the OEM is responsible for the accessibility.

**SP.** As mentioned in relation to **Q5.1**, Company X shall involve a wide range advocacy groups in development of the data management strategy. Example stakeholder representation include elderly, wheelchair users, ethnic minorities, and children.

*F. [REQ6] Societal and environmental well-being*

**[REQ6]** covers **Environmental Well-Being** (**EWB Q6.1-6.2.1**), **Impact on Work and Skills** (**IWS Q6.3-6.7.1**), and **Impact on Society at Large or Democracy** (**ISLD Q6.8-6.8.3**).

*1) ALTAI Applicability for ADAS*

**[REQ6]** considers the big picture of societal and environmental impact of AI systems. An ADAS has only marginal and indirect impact on such high-level considerations. One could argue that ADAS development contributes to the future introduction of large-scale Autonomous Driving (AD). Many predict that AD will have a substantial impact on society, e.g., disrupting the real estate industry and increasing truck driver unemployment. Others speculate that AD will help reaching the UN sustainability goals, e.g., supporting mobility-as-a-service resulting in fewer vehicles in operation and a significant decrease in traffic accidents. We argue, however, that all the **[REQ6]** Qs are N/A, both for SMIRK and ADAS development in general.

*G. [REQ7] Accountability*

**[REQ7]** contains two subsections: **Auditability (Q7.1-7.2)** and **Risk Management** (**RM Q7.3-7.8**).

*1) ALTAI Applicability for ADAS*

**[REQ7]** covers auditability and risk management, two aspects that are highly important in ADAS development. We consider all Qs applicable, except **Q7.8**. The driver remains ultimately responsible in a car supported by ADAS, thus there is no need to implement redress by design mechanisms. While the **[REQ7]** Qs are relevant to ADAS, we find them surprisingly redundant – most Qs overlap with **[REQ2]** or **[REQ4]**.

First, the auditability subsection contains two questions that were covered by the traceability and explainability subsection of **[REQ4]**, i.e., the key requirement on transparency. This part of ALTAI brings no additional value, except the clarification of the third-party perspective. The RM subsection largely overlaps with the **[REQ2]** subsection of general safety. Risk management is the backbone of safety engineering. For the next revision of ALTAI, we strongly recommend merging auditability with **[REQ4]** and moving **Q7.4**, **Q7.7**, and **Q7.7.1** to **[REQ2]**. After this move, we propose renaming the RM subsection to "Ethics and Process Management" to better reflect its content.

*2) SMIRK ALTAI Assessment*

**Auditability.** Auditability is essential to the development of SMIRK, as Company X will use external auditors to assess the safety evidence required for ISO 26262 certification (**Q7.1-7.2**).

**RM.** Company X does not have an internal AI ethics board (**Q7.5**), but closely monitors discussions in the automotive domain. As ALTAI is not a one-shot assessment, Company X has set up a process to repeat the process similar to other SMIRK re-certification efforts (**Q7.6**).

## VI. Summary and Concluding Remarks

We presented an illustrative case study using ALTAI for a fictive Tier 1 automotive supplier developing the ML-based SMIRK ADAS. First, we explored how applicable ALTAI is for ADAS development (RQ1). Second, we discussed how SMIRK must be developed to comply with ALTAI (RQ2), i.e., how to ensure that SMIRK qualifies as a trustworthy AI system.

We found that ALTAI is largely applicable to ADAS development. Certain parts related to human agency and transparency are N/A as an ADAS primarily interacts with the ego car rather than the human driver. Moreover, parts related to larger societal and environmental issues are too broad for an ADAS supplier. The evaluative scope for ADAS should be limited to its context.

Our fictive company exercise suggested that an ALTAI-compliant SMIRK ADAS is possible. The major work is connected to technical robustness and safety, but well in line with existing and emerging automotive standards. Appropriate safety and security certification would be sufficient for ALTAI. Several ALTAI questions must be tackled together with the OEMs and other ADAS suppliers, including how to propagate information to the driver if dozens of ADAS are in use and how to combat overreliance on automated driving. Finally, data governance is a major concern: we assumed SMIRK would collect its own data, but future OEMs might prefer to orchestrate all data. We conclude with three recommendations for a revised ALTAI.

First, ALTAI should support organizations in different life-cycle stages. Just like other fundamental quality attributes, ethics cannot be added on top of an existing system. Concepts such as quality-by-design are popular, but ALTAI often appears like checklists for a complete product. We recommend creating ALTAI variants for separate development phases, e.g., conceptual, development, and operations. Moreover, contemporary software and AI engineering are done in highly iterative fashion, resulting in practices such as continuous-X and DevOps. ALTAI must facilitate reassessments, analogous to initiatives to support dynamic safety cases and recertification in safety engineering.

Second, domain-specific adaptations of ALTAI should evolve. Development organizations would find ALTAI more actionable with more questions tailored to the context. The general IEC 61508 safety standard has branched to different domains and ALTAI should follow suit. We recommend appointing taskforces of ethicists and engineers active in various domains, e.g., MedTech, FinTech, and automotive. As a first step, the task forces could add domain-specific examples to the existing ALTAI. For example, when assessing the accountability of an ADAS supplier, it became obvious that developers of AI systems in complex ecosystems need further guidance, in line with Floridi *et al.* [1].

Third, ALTAI should be reorganized to support readability. While we understand that ALTAI is based on the seven key requirements of trustworthy AI, the current structure hinders ALTAI applicability. The sections in the current ALTAI revision are unbalanced, i.e., [REQ2] covers too much and should be split. Moreover, [REQ7] is partly redundance as it overlaps with [REQ2] and [REQ4] – unexpected redundancy is surprising. Finally, all checklist items shall, of course, have unique identifiers.

## Acknowledgements

This work was funded by Kompetensfonden at Cam-pus Helsingborg, Lund University, Sweden. Furthermore, the project received financial support from the SMILE III project financed by Vinnova, FFI, Fordonsstrategisk forskning och innovation under the grant number: 2019-05871 and the EC-SEL Joint Undertaking (JU) under grant agreement No 876852(VALU3S).